\documentclass[aps,prl,twocolumn,showpacs]{revtex4}
\usepackage{psfrag}
\usepackage{epsfig}
\usepackage{color}

\newcommand     {\beq}[1]         { \begin{equation} #1 \end{equation} }

\begin{document}

\title{Damage and Healing in Fatigue Fracture}
\author{F.\ Kun${}^1\footnote{Electronic 
address:feri@dtp.atomki.hu}$, M.\ H.\ A.\ S.\ Costa${}^{2}$, R.\ N.\
Costa Filho${}^{2}$, J.\ S.\ Andrade Jr.${}^{2}$, J.\ B.\
Soares${}^{3}$, \\ S.\ Zapperi${}^{4}$, and H.\ J.\ Herrmann${}^{5}$} 
\affiliation{ 
\centerline{${}^1$Department of Theoretical Physics, University of Debrecen, P.\
O.\ Box:5, H-4010 Debrecen, Hungary}  \\
\centerline{${}^2$ Departamento de F\'{\i}sica, Universidade Federal
do Cear\'a, 60451-970 Fortaleza, Cear\'a, Brazil} \\
\centerline{${}^3$ LMP, DET, Universidade Federal
do Cear\'a, 60451-970 Fortaleza, Cear\'a, Brazil} \\
\centerline{${}^4$ CNR-INFM, Dipartimento di Fisica, Universit\'a
'La Sapienza', Piazzale Aldo Moro 2, 00185 Roma, Italy} \\
${}^5$ IfB, HIF, E12, ETH, Hoenggerberg, 8093 Z\"urich, Switzerland}
\date{\today}
\begin{abstract}
We present an experimental and theoretical study of the fatigue
failure of asphalt under cyclic compression. Varying the load
amplitude, experiments reveal a finite fatigue limit below which  
the specimen does not break, while approaching the tensile strength of
the material a rapid failure occurs. In the intermediate load range,
the lifetime decreases with the load as a power law. We 
introduce two novel theoretical approaches, namely, a fiber bundle
model and a fuse model, and show that both capture the major
microscopic mechanisms of the fatigue failure of 
asphalt, providing an excellent agreement with the experimental
findings. Model calculations show that the competition of
damage accumulation and healing of microcracks gives rise to novel
scaling laws for fatigue failure.
\end{abstract}
\pacs{62.20.Mk, 46.50.+a, 61.82.Pv}
\maketitle

The fracture of disordered media
represents an important applied problem, with intriguing
theoretical aspects. Statistical models have been 
successfully applied in the past to analyze fracture under
quasistatic conditions, but the effect of  cyclic loading
is less explored \cite{review}. Laboratory experiments
reveal that fatigue failure under repeated
loading is due to a combination of several mechanisms, 
among which damage growth, relaxation due to 
viscoelasticity, and healing of microcracks play an essential role
\cite{yongki_asph_2002,little_healing_2002_1,sornette_fatigue_1992}.
Theoretical approaches have serious difficulties to capture all these
mechanisms
\cite{sornette_fuse_fatigue,little_healing_2002_1,chakrabarti_fatigue,sornette_fatigue_1992,lee_kim_1998_jem,yongki_asph_2002}
and fatigue life prediction is still very much an 
empirical science. Understanding this problem has crucial implications
even for everyday applications. For example, fatigue failure occurring
in roads due to repeated traffic loading cause main distress,
limiting the lifetime of asphalt pavements.

\begin{figure}
  \begin{center}
\psfrag{aa}{\textcolor{white}{\LARGE \bf $\sigma_0$}}
\psfrag{bb}{\large $b)$}
\psfrag{cc}{\large $a)$}
\epsfig{bbllx=0,bblly=0,bburx=315,bbury=135,file=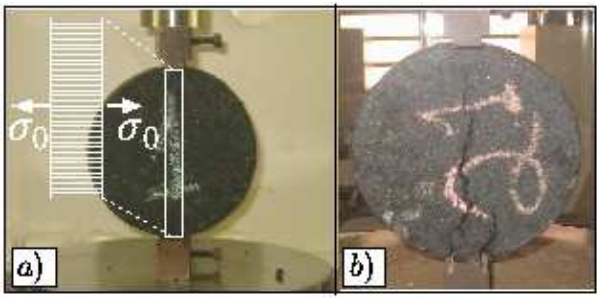,
  width=8.0cm}
   \caption{
Setup of the experiments. $a)$ A cylindrical asphalt sample
is subjected to diametrical compression applied
periodically. FBM discretizes the region where tensile
stress emerges (white rectangle) in terms of fibers. 
$b)$ At complete failure a crack spans the cylinder along
the load direction.}
\label{fig:exper_setup}
  \end{center}
\end{figure}

In this Letter we present a detailed experimental and theoretical
study of the fatigue performance of hot mix asphalt (HMA).  
We carried out 
fatigue life tests of  specimens measuring the accumulation of 
deformation with the number of loading cycles and the lifetime of
specimens varying the load amplitude. To obtain a theoretical
understanding of the experimental findings, we worked out two novel
modelling approaches for fatigue failure, namely, a fiber bundle
model \cite{sornette_prl_78_2140,kun_epjb_17_269_2000}
and a fuse model \cite{dearcangelis85}. We then show that
both descriptions capture the stochastic nature of the fracture process, the 
immediate breaking of material elements and the cumulative effect of
the loading history. Two physical mechanisms are considered which
limit the accumulation of damage: a finite activation threshold of
crack nucleation below which the local load does not
contribute to the ageing of the material and 
 healing of microcracks under compression,
which leads to damage recovery. The analytical and
numerical results of the model calculations provide a good
quantitative agreement with the experimental findings. 
We show that the competition of nucleation
and healing of microcracks leads to novel type of scaling laws for
fatigue fracture with universal scaling exponents. 

\begin{figure}
  \begin{center}
\psfrag{aa}{\large $a)$}
\psfrag{bb}{\large $b)$}
\epsfig{bbllx=110,bblly=300,bburx=450,bbury=580,file=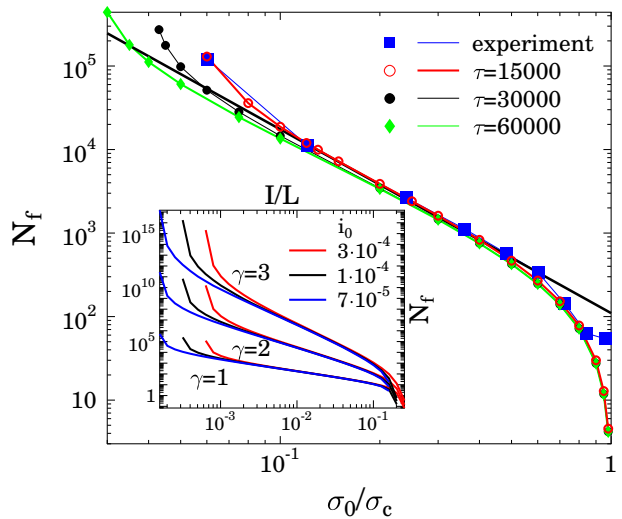,
  width=8.2cm}
   \caption{{\it (Color online)} $N_{f}$ as function of load
$\sigma_0/\sigma_c$ for FBM 
varying the value of $\tau$. FBM provides an excellent fit of the  
experimental results with $\gamma=2.0$, $\tau=15000$,
and $a=0.01$. Fuse model results are presented in the inset varying
the value of $\gamma$ and the threshold current $i_0$.
 }
\label{fig:n_cycle}
  \end{center}
\vspace*{-0.5cm}
\end{figure}

In order to obtain a quantitative characterization of the process of
fatigue failure, we carried out fatigue life tests of asphalt under cyclic
diametric compression of cylindrical specimens at a constant external
load $\sigma_0$ (see Fig.\ \ref{fig:exper_setup}). 
HMA is the primary material
used to construct and maintain pavements and roadways due to its good
mechanical performance and high durability. From the structural point of
view asphalt is a combination of aggregates (usually crushed stone and
sand), filler (cement, hydrated lime or stone dust) and a
bituminous binder. 
Cylindrical samples of HMA were produced using the Marshall method
\cite{marshal_kandhal} and then loaded by a hydraulic device. 
Under repeated loading at a constant amplitude
$\sigma_0$, the deformation 
$\varepsilon$ was monitored as a function of the number of cycles
$N_{cycle}$. Furthermore, the total number of cycles to complete
failure $N_{f}$ was measured varying  $\sigma_0$.   
\begin{figure}
  \begin{center}
\epsfig{bbllx=20,bblly=10,bburx=480,bbury=400,file=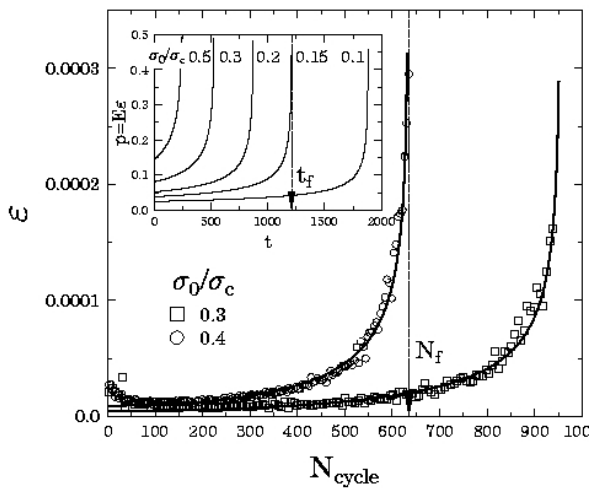,
  width=7.5cm}
   \caption{Deformation as a function of the number of
loading cycles. The continuous lines result from our theory. Inset:
Load on the fibers in FBM as function of time  at different values 
of $\sigma_0/\sigma_c$ for uniformly distributed threshold values setting
$\tau = \infty$ in Eq.\ (\ref{eq:eom}).
 }
\label{fig:p_t}
  \end{center}
\end{figure}
Figure \ref{fig:p_t} presents representative examples of
$\varepsilon(N_{cycle})$ recorded at loads $30 \%$ and $40 \%$ of
the tensile strength $\sigma_c$ of the specimen. It can be observed
that due to the gradual accumulation of damage, the deformation
$\varepsilon$ caused by the same load $\sigma_0$ monotonically
increases until catastrophic failure occurs after a finite number of
cycles $N_{f}$. The 
derivative  of $\varepsilon(N_{cycle})$ also shows a monotonous
increase and diverges when approaching the point of
macroscopic failure.  
Increasing the external load the functional form of
$\varepsilon(N_{cycle})$ remains the same, however, the lifetime of the
specimen $N_{f}$ gets shorter. 
The fatigue lifetime $N_{f}$ measured at different fractions of
the tensile strength $\sigma_c$ (Fig.\
\ref{fig:n_cycle}) reveals the existence of three distinct
regimes. First, approaching the
tensile strength of the material $\sigma_0/\sigma_c \to 1$ the
lifetime $N_f$ rapidly decreases indicating an immediate 
failure of the specimen. At the other extreme, a lower
threshold value of the external load $\sigma_l$ can be identified
below which the specimen suffers only partial damage giving rise to
an infinite lifetime (fatigue limit). In the intermediate regime 
the experimental results follow a power law known as
Basquin law 
\cite{basquin_1910,yongki_asph_2002,little_healing_2002_1,sornette_fatigue_1992}
\begin{eqnarray}
\label{eq:basq}
N_{f} \sim \left(\frac{\sigma_0}{\sigma_c}\right)^{-\alpha},
\end{eqnarray} 
where $\alpha = 2.2 \pm 0.1$, as shown in Fig.\
\ref{fig:n_cycle}. 

The experiments show that the fatigue
crack growth  is localized
to a narrow region between the loading plates (see
Fig.\ \ref{fig:exper_setup}$b$) where locally a tensile
stress emerges perpendicular to the external load
\cite{turcotte_dam_reol_fbm}. 
To give a theoretical description of the failure process, we focus on
this region and discretize it by a fiber bundle model (FBM) as
illustrated in Fig.\ \ref{fig:exper_setup}$a$
\cite{sornette_prl_78_2140,kun_epjb_17_269_2000}. 
We consider a bundle of parallel linear elastic
fibers with the same 
Young modulus $E$. Under diametrical compression of the disc-shaped
specimen, the fibers experience a tensile loading and gradually fail
due to immediate breaking or to
the ageing of material elements
\cite{yongki_asph_2002}. More precisely, the following two mechanisms
are considered: $(I)$ fiber 
$i$ ($i=1, \ldots , N)$ breaks instantaneously at time $t$ when its
local load $p_i(t)$ exceeds the tensile strength $p_{th}^i$ of the
fiber. $(II)$ All intact fibers undergo a 
damage accumulation process due to the load they have experienced.
The amount of damage
$\Delta c_i$ occurred under the load $p_i(t)$ in a time
interval $\Delta t$ 
is assumed to have the form $\Delta c_i = a
p_i(t)^{\gamma}\Delta t $ \cite{turcotte_dam_reol_fbm}, hence, the
total accumulated damage $c_i(t)$ until 
time $t$ can be obtained by integrating over the entire loading
history \cite{turcotte_dam_reol_fbm}. The exponent $\gamma>0$ controls
the damage accumulation rate and $a>0$ is a scale parameter. 
The fibers can only tolerate a
finite amount of damage and break when $c_i(t)$ exceeds a threshold
value $c_{th}^i$.
The two breaking thresholds $p_{th}^i$ and $c_{th}^i$ are random
variables with a joint probability density function
$h(p_{th},c_{th})$. Assuming independence 
of the two breaking modes, $h$ can be factorized into a product 
$h(p_{th},c_{th}) = f(c_{th}) g(p_{th})$, where $f(c_{th})$ and
$g(p_{th})$ are the probability densities and  $F(c_{th})$ and
$G(p_{th})$ the cumulative distributions of the breaking thresholds
$p_{th}$ and $c_{th}$, 
respectively.  For simplicity, after each breaking event the load of
the broken fiber is equally redistributed over the intact ones
irrespective of their distance from the failure point (global load
sharing) \cite{sornette_prl_78_2140,kun_epjb_17_269_2000}. 

Under a constant tensile load $\sigma_0$, the load on a single fiber
$p_0$ is initially determined by the quasi-static constitutive
equation of FBM $\sigma_0 = \left[1-G(p_0) \right]p_0$
\cite{sornette_prl_78_2140,kun_epjb_17_269_2000}.  
The external load $\sigma_0$ must fall below the tensile strength of
the bundle $\sigma_0 < \sigma_c$, otherwise the bundle will fail immediately.
\begin{figure}
\psfrag{aa}{$a)$}
\psfrag{bb}{$b)$}
  \begin{center}
\epsfig{bbllx=0,bblly=0,bburx=700,bbury=320,file=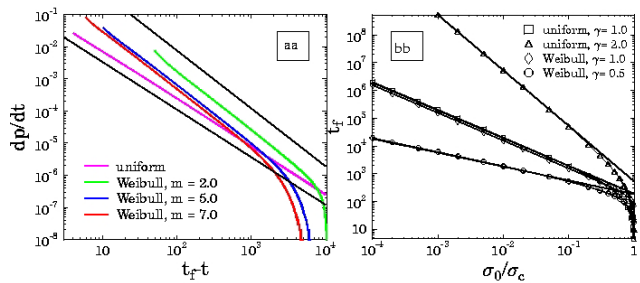,
  width=8.4cm}
   \caption{{\it (Color online)} $a)$ Deformation rate $dp/dt$ as
function of the time to macroscopic failure for
different disorder distributions and $\gamma = 1$. Straight lines are
drawn to guide the eye with slopes 1.5 and 1.8. $b)$ $t_f$ as a
function of $\sigma_0/\sigma_c$ for uniform and Weibull 
distributions ($m=2.0$) varying the value of $\gamma$. The slope of the
straight lines is equal to the  value of the corresponding
exponent $\gamma$.
}
\label{fig:derivative}
  \end{center}
\end{figure}
As time elapses, the fibers accumulate damage and break due to their
finite damage tolerance. These breakings, 
however, increase the load on the remaining intact fibers 
which in turn induce again immediate breakings.
This way, in spite of the independence of the threshold values
$p_{th}$ and $c_{th}$, the two breaking modes are dynamically
coupled, gradually driving the system to macroscopic failure in a
finite time $t_f$ at any load values $\sigma_0$. 
Healing of microcracks can be captured in the model by introducing a
finite range $\tau$ for the memory, over which the loading history
contributes to the accumulated damage
\cite{little_healing_2002_1,hans_memory_86}.   
Finally, the evolution equation of the system can be cast in the form 
\beq{
\sigma_0 = [1-F(a \int \limits_0^t e^{-\frac{(t-t')}{\tau}}
p(t')^{\gamma}dt')]\left[1-G(p(t))\right]p(t), 
\label{eq:eom}
}
where the integral in the argument of $F$ provides the accumulated
damaged at time $t$ taking into account the finite range of memory by
the exponential term \cite{hans_memory_86}.  In principle, the range of memory $\tau$
can take any positive value $\tau > 0$ such that during the time evolution of
the bundle the damage accumulated during the time interval $t' <
t-\tau$ heals. Equation (\ref{eq:eom}) is an integral equation
which has to be solved for the load $p(t)$ on the intact fibers at a
given external load $\sigma_0$ with the initial condition $p(t=0) =
p_0$. The product in Eq.\ (\ref{eq:eom}) arises due to the
independence of the two breaking thresholds. We
note that Eq.\ (\ref{eq:eom}) recovers the usual constitutive
behavior of FBM \cite{sornette_prl_78_2140,kun_epjb_17_269_2000}
when damage accumulation is suppressed either by
increasing the exponent $\gamma$ or decreasing the range of memory
$\tau \to 0$. 

The inset of Fig.\ \ref{fig:p_t} presents examples of the solution
$p(t)$ of Eq.\ (\ref{eq:eom}) obtained for breaking thresholds uniformly
distributed  in the interval $[0,1]$ at different ratios
$\sigma_0/\sigma_c$
setting $\tau \to \infty$. Since $p(t)$ is simply related to the
macroscopic deformation $\varepsilon$ of the bundle $p(t) =
E\varepsilon (t)$, these results can directly be compared to
the experimental findings. 
The agreement seen in Fig.\ \ref{fig:p_t} is obtained using Weibull
distributions $P(x) = 1-\exp{\left[-\left(x/\lambda\right)^m\right]}$
for the two breaking thresholds in Eq.\ (\ref{eq:eom}) with the same
Weibull exponent $m$ and scale parameter $\lambda$.  
Our calculations also reveal that approaching macroscopic
failure, the derivative of $p(t)$ has a power law divergence as a
function of the time to failure
$\frac{dp}{dt} \sim (t_f-t)^{-\beta}$.
Fig.\ \ref{fig:derivative}$a$ shows that the exponent
$\beta$ solely depends on the type of 
disorder. Specifically, for breaking thresholds distributed over a finite
and infinite range, we obtain the exponents $\beta = 1.5 \pm 0.02$ and
$\beta = 1.8 \pm 0.06$, respectively, defining two different
universality classes of the fatigue failure.

It is possible to recover the Basquin law Eq.\ (\ref{eq:basq}) from 
Eq.\ (\ref{eq:eom}), {\it i.e.} it can be shown analytically 
that for $\sigma_0/\sigma_c << 1$ and $\tau \to \infty$  the lifetime
of the system has a power law dependence on the external load 
$t_f \sim \left(\frac{\sigma_0}{\sigma_c}\right)^{-\gamma}$,
where  $\gamma$ is the damage accumulation
exponent, independent on the type of disorder. 
Figure \ref{fig:derivative}$b$ shows that the numerical
results are in excellent agreement with the above analytic
prediction. 

Due to Eq.\ (\ref{eq:eom}), without
healing ($\tau \to \infty$) the cumulative effect of the loading history 
gives rise to a macroscopic failure of the system at any load.
However, our experiments revealed that damage 
recovery caused by healing of microcracks results in a finite fatigue
limit $\sigma_l$, below which the sample does
not break. Since healing takes place in the polymer binder, it
can be controlled by changing the temperature
\cite{little_healing_2002_1}.
\begin{figure}
\psfrag{aa}{{\LARGE $a)$}}
  \begin{center}
\epsfig{bbllx=145,bblly=450,bburx=480,bbury=720,file=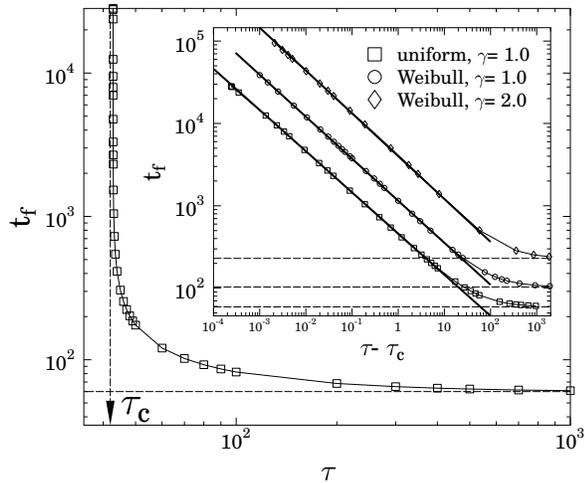,
  width=7.8cm}
   \caption{Lifetime $t_f$ of FBM at a constant
load $\sigma_0/\sigma_c = 0.7$ varying $\tau$. Approaching $\tau_c$ the lifetime
diverges. Inset: $t_f$ as a function of the distance from the critical
range of memory $\tau - \tau_c$ for different disorder distributions
and $\gamma$ exponents. The horizontal lines indicate the value of
$t_f(\tau=\infty)$. 
}
\label{fig:tau_time}
  \end{center}
\end{figure}
In our FBM the healing of microcracks is captured by the finite range
of memory $\tau$. This is illustrated in Fig.\ \ref{fig:tau_time}
presenting the lifetime $t_f$ for a fixed load varying the value of
$\tau$ over a broad range.  
It can be seen that by decreasing $\tau$, the lifetime of the system
increases and, due to the 
competition between nucleation of new microcracks and healing of the
existing ones, a finite critical value $\tau_c$ emerges below
which the system only suffers a partial damage and has an infinite
lifetime $t_f \rightarrow \infty$. Plotting the lifetime $t_f$ as
function of 
the distance from the critical point $\tau_c$ (inset of Fig.\
\ref{fig:tau_time}) we find a power law dependence of $t_f$ 
\beq{
t_f \sim (\tau - \tau_c)^{-\delta},
}
where $\delta = 0.5 \pm 0.01$ is an universal exponent,
independent of the type of disorder and of the damage accumulation
exponent $\gamma$. Consequently, at a given temperature
where the system is characterized by a fixed value of $\tau$, a
finite fatigue limit $\sigma_l$ emerges at which (Fig.\
\ref{fig:n_cycle}) the lifetime diverges
\cite{little_healing_2002_1}.

Besides healing, another important mechanism which limits damage
accumulation is a finite activation threshold of microcrack
nucleation. In order to study this effect we consider the random fuse
model (RFM) \cite{dearcangelis85} 
of fracture and extend it by introducing a history
dependent ageing variable of fuses. We construct an $L\times L$ tilted
square lattice of initially fully intact bonds with identical 
conductance but random failure thresholds $i_c$. The threshold values
$i_c$ are uniformly distributed between a small current value $i_0$
and 1 ($i_0 << 1$). For a given value of current $I$ applied between
two bus bars of the lattice, the local current through each bond is
determined by solving numerically the Kirchhoff equations. Fuses
burn out irreversibly when the current exceeds the local failure
thresholds. This process is then followed by the recalculation of the
current values. In order to capture fatigue cracking in the model, intact
fuses are assumed to undergo an ageing process, modeled 
by a variable $A(t) =
\sum_{t'=1}^t a(i(t')-i_0)^{\gamma}$. 
A fuse fails due to fatigue when $A(t)>A_{max}$, where
$A_{max}$ is a failure threshold uniformly distributed between $1-b$
and $1+b$ with $b=0.1$. Comparing the two modelling
approaches, the ageing variable $A(t)$ of RFM is analogous to the
accumulated damage $c(t)$ of FBM, however, only current values above
$i_0$ contribute to $A(t)$, which captures the finite activation
threshold of microcrack nucleation. The inset of Fig.\ 
\ref{fig:n_cycle} demonstrates that RFM of ageing fuses provides
qualitatively the same behavior as FBM, {\it i.e.} rapid failure at
high current values $I$, a Basquin regime Eq.\ (\ref{eq:basq}) at
intermediate currents with an exponent equal to $\gamma$ and a finite
fatigue limit $\sigma_l$ determined by the threshold current $i_0$. 

In summary, our experiments on the fatigue failure of
asphalt under cyclic compression revealed three regimes of the failure
process depending on the load amplitude: instantaneous breaking, a
Basquin regime of a power law decrease of lifetime and the existence
of a fatigue limit below which no failure occurs. We introduced two
novel modeling approaches both
capturing the essential ingredients of the fatigue failure of bituminous
materials. These
models provide a comprehensive description of the experimental
findings and additionally revealed novel scaling laws of fatigue
fracture: approaching the macroscopic failure, the process of fatigue
fracture accelerates and is characterized by a finite time power law
singularity of the deformation rate. The exponent is different for
bounded and unbounded disorder distributions defining two universality
classes of fatigue fracture. Due to the interplay between damage and
healing, at each load level a critical value of
the range of memory 
emerges where the lifetime of the system has a universal power law
divergence. 

We thank the Brazilian agencies CNPq, CAPES, FUNCAP and FINEP for
financial support. F.\ Kun was supported by NKFP-3A/043/04
and OTKA T049209. H.\ J.\ Herrmann is  
grateful for the Max Planck Prize.

\end{document}